\documentclass[aps, pra, singlecolumn, showpacs, a4paper, nofootinbib, superscriptaddress,longbibliography, floatfix]{revtex4-2}

\usepackage{graphicx}
\usepackage{dcolumn}
\usepackage{bm}
\usepackage{color, xcolor}
\usepackage{natbib}
\usepackage{amssymb,amsfonts,amsmath,mathtext,longtable,array}
\usepackage{appendix}
\usepackage{multirow}
\usepackage{svg}

\begin{document}

\title{Radio-Frequency  Hong-Ou-Mandel Interference {with Conditionally Built States}}

\author{A. Sheleg}
\affiliation{School of Electrical and Computer Engineering, Tel Aviv University, Tel Aviv 69978, Israel}

\author{D. Vovchuk}
\affiliation{School of Electrical and Computer Engineering, Tel Aviv University, Tel Aviv 69978, Israel}

\author{K. Boiko}
\affiliation{School of Electrical and Computer Engineering, Tel Aviv University, Tel Aviv 69978, Israel}

\author{P. Ginzburg}
\affiliation{School of Electrical and Computer Engineering, Tel Aviv University, Tel Aviv 69978, Israel}

\author{G. Slepyan}
\affiliation{School of Electrical and Computer Engineering, Tel Aviv University, Tel Aviv 69978, Israel}

\author{A. Boag}
\affiliation{School of Electrical and Computer Engineering, Tel Aviv University, Tel Aviv 69978, Israel}

\author{A. Mikhalychev}
\affiliation{Atomicus GmbH, Amalienbadstr. 41C, Karlsruhe, 76227 Germany}

\author{A. Ulyanenkov}
\affiliation{Atomicus GmbH, Amalienbadstr. 41C, Karlsruhe, 76227 Germany}

\author{T. Salgals}
\affiliation{Institute of Photonics, Electronics and Telecommunications, Riga Technical University, Azenes str. 12 – 234, Riga, LV – 1048, Latvia }

\author{P. Kuzhir}
\affiliation{Department of Physics and Mathematics, University of Eastern Finland, Yliopistokatu 7,FI-80101 Joensuu, Finland}

\author{D. Mogilevtsev}
\affiliation{School of Electrical and Computer Engineering, Tel Aviv University, Tel Aviv 69978, Israel}

\begin{abstract}
We report {an experimental demonstration of room-temperature Hong-Ou-Mandel (HOM) interference at a radio-wave frequency of 120 MHz using  conditional build-up of quantum states from classical phase-averaged coherent states.} This approach enables {observation} of quantum effects in spectral regimes where conventional single-photon sources and detectors are unavailable or require cryogenic conditions. By constructing a high-fidelity approximation of a single-photon state with phase-averaged coherent states, we observe the normalized second-order intensity correlation dips significantly below the classical limit of 0.5. The method allows for tunable noise suppression via optimization of the state representation. Our results establish the feasibility of using conditionally prepared classical states to simulate quantum interference phenomena in the radio-frequency domain. This technique opens the door to realizing other quantum protocols, such as Bell inequality tests, in frequency ranges where standard quantum technologies are currently infeasible.

\end{abstract}


\maketitle

\section{Introduction}

One of the main cornerstones of modern quantum photonic technologies in any given frequency range is availability of sources producing quantum states and detecting devices able to capture effects of state quantumness in this wavelength region. Spectacular recent advances in photonic quantum sensing  and metrology are reached  in the visible and near-infrared frequency ranges, where such sources are readily available \cite{RevModPhys.89.035002,10.1116/5.0007577,10.1117/1.OE.53.8.081910}. In particular, single photons can be produced by a variety of sources in these frequency ranges, for instance,  by single quantum emitters such as atoms \cite{doi:10.1126/science.1113394,PhysRevLett.89.067901}, or molecules  \cite{PhysRevLett.76.900,PhysRevLett.83.2722,lounis},  
 color centers in diamonds \cite{babi,PhysRevLett.110.243602,Naydenov2015}, or semiconductor quantum dots \cite{PhysRevLett.86.1502,2017NatNa..12.1026S}.  Single photons in these wavelength ranges can be produced by heralding photons from the source that creates correlated photons pairs, for example, by very widely used spontaneous down-conversion process \cite{PhysRevLett.25.84}.   
 Recently, down-conversion sources of twin-photons were realized in the microwave frequency range with the help of super-conducting Josephson junctions \cite{houk,2013NatPh...9..345L}.  However, their engineering applications remain limited by the required cryogenic temperatures. Also, there are still electromagnetic spectral regions where quantum states' generators and/or single-photon detectors are hardly available. For example, there is so called "teraherz gap" in the region $5-15$ THz,  where quantum state generators are still in the early experimental stage \cite{TodorovDhillonMangeney+2024+1681+1691}. Furthermore, quantum effects for the wavelengths longer than in the microwave range are practically unavailable.

Very recently, a method was suggested that allows a conditional build-up of quantum states from the classical states \cite{PhysRevA.105.052206,https://doi.org/10.1002/qute.202300060}.
 This method exploits information about quantum states structure for reproducing the results of measurements for some observables (in particular, normally ordered correlation field functions). {At the price of increased statistical errors and/or measurement time,} it allows reproducing quantum effects in the wavelength regions where quantum state generators and single-photon detectors are unavailable or overly expensive and require specific conditions such as cryogenic operation (as it is for microwave quantum photonics).
 
 Here, we report a first experimental implementation of such a fundamental quantum effect as the Hong-Ou-Mandel (HOM) interference at a radio-wave frequency of 120 MHz (wavelength of 2.5 m) based on this method. 
Originally devised to measure picosecond-scale time intervals between single-photon pulses, the HOM interference experiment has since become a cornerstone demonstration of the quantum nature of the electromagnetic field, exhibiting behavior with no classical analog \cite{PhysRevLett.59.2044}. For a comprehensive overview of its development and diverse applications, see Ref. \cite{Bouchard_2021}. Beyond its foundational significance, the HOM effect plays a central role in practical quantum technologies, including quantum communication, computation, and metrology \cite{2023NatRP...5..326C}. It is employed to verify the indistinguishability of single-photon states \cite{2009NatPh...5..715B,Ornelas-Huerta:20}, implement destructive SWAP tests \cite{PhysRevA.87.052330}, and enhance quantum sensing protocols \cite{Cassemiro_2010,Basiri-Esfahani:15}, as well as in architectures for linear optical quantum computing \cite{Nielsen_Chuang_2010,2001Natur.409...46K,RevModPhys.79.135}.

The uniquely quantum nature of two-particle interference at the heart of the HOM effect has been observed not only with photons, but also with other bosonic and fermionic particles, including phonons \cite{PhysRevLett.117.180501}, electrons \cite{doi:10.1126/science.1232572}, atoms \cite{doi:10.1126/science.1250057}, and plasmons \cite{128}. HOM interference using optical and near-infrared photons has become sufficiently well-established to serve as a standard laboratory experiment in undergraduate quantum optics courses \cite{Carvioto-Lagos_2012,10.1119/5.0210869}.

Nonetheless, there remain regions of the electromagnetic spectrum where HOM interference has yet to be realized. Expanding the reach of HOM-type experiments into these spectral regimes could offer new avenues for both foundational studies and emerging quantum technologies.
By employing phase-averaged coherent states to conditionally construct scaled single-photon-like states, here we observe a HOM interference dip that falls significantly below the classical limit of 0.5 for the normalized second-order intensity correlation function between two detectors.   This experiment supports the theoretical prediction \cite{PhysRevA.105.052206,https://doi.org/10.1002/qute.202300060} and opens {the way of observing quantum effects at} radio-wave frequencies much lower than microwaves.

The outline of the paper is as follows. In Section II, we briefly explain the method of conditionally building up quantum states as applied for our case. In Section III, the experimental setup and the experiment are described. The fourth section reports experimental results for different  state build-up strategies. The HOM dips for classical and quantum states are shown. Finally, concluding remarks and observations are provided in Section V.

\begin{figure}
    \centering
    \includegraphics[width=0.99\linewidth]{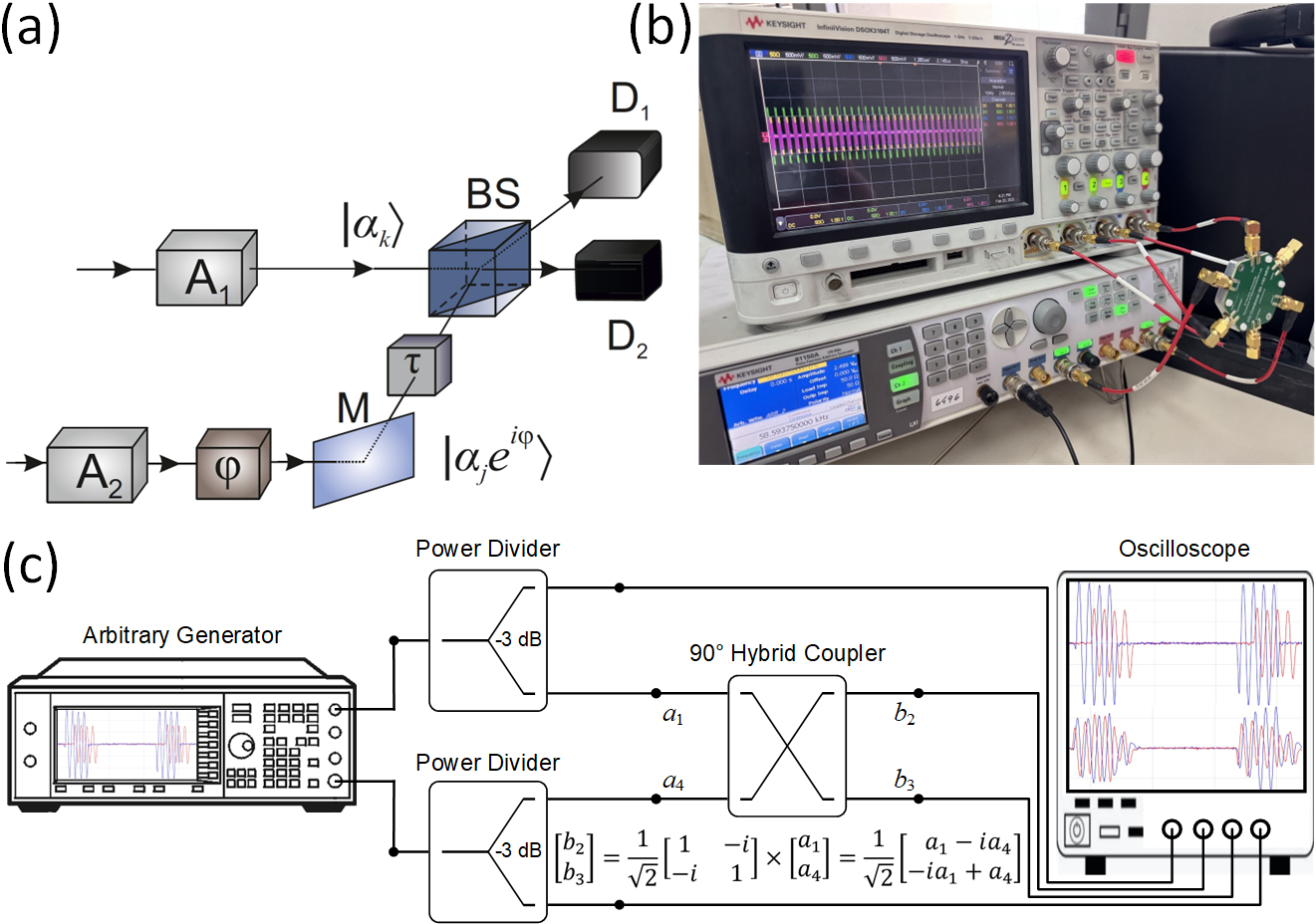}
    \caption{ (a) Schematic outline of the HOM interference experiment. (b) Photograph of the actual experimental setup. (c) Block-scheme of the setup. }
    \label{fig1}
\end{figure}

\begin{figure}
    \centering
    \includegraphics[width=0.99\linewidth]{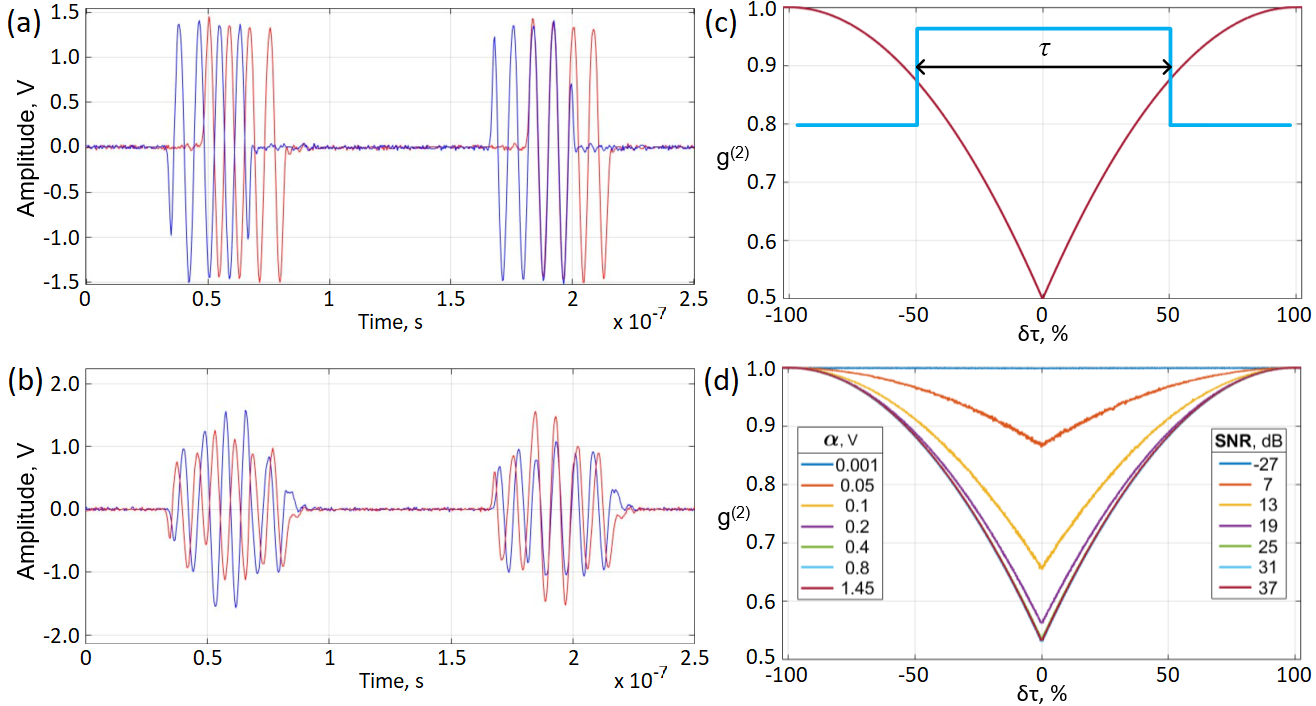}
    \caption{ (a) Illustration of partially overlapping radio pulses of the same amplitude, which were used in the experiment. (b) An illustration of the completely overlapping radio pulses after the beam-splitter. (c) An illustration of the HOM dip for the interference of identical phase-averaged coherent states carried by rectangular radio pulses. The dip is shown versus  the relative time-shift of the pulses expressed as the percentage of the pulse width. (d) HOM dips obtained for the realistic pairs of radio pulses of the same shape and different amplitudes proportional to the respective voltages shown in the left inset; the right inset shows the signal-to-noise ratio (SNR) for the corresponding HOM curves. {Notice that for the curves corresponding to $\alpha=0.4$, $\alpha=0.8$, and $\alpha=1.45$ the SNR exceeds 25 dB and these three curves are almost completely overlapping.}}
    \label{fig2}
\end{figure}

\section{Building up quantum states}

The method of conditionally building up quantum states stems from the ``data pattern" approach to quantum tomography, when the density matrix of an unknown signal state $\rho_s$ is fitted with a  non-convex  combination of the density matrices $\rho_j$ of classical probes  \cite{PhysRevLett.105.010402}:
\begin{equation}
\rho_s\approx \sum\limits_{\forall j} c_j\rho_j,
\label{datapattern}
\end{equation}
where the coefficients $c_j$ are real, but can be negative \footnote{Notice that the fitting procedure is generally nonlinear by providing for non-negativity of the result.}.  
The method of the conditional building up quantum states is actually reversing Eq. (\ref{datapattern}) \cite{PhysRevA.105.052206,https://doi.org/10.1002/qute.202300060}. One takes a mixture of classical states with corresponding coefficients $c_j$ for reproducing results of a measurement over a quantum states $\rho_s$. Negativity of the coefficients $c_j$  can be achieved in practice by adding to the set $\rho_j$ a binary ancilla encoding signs. So, to reproduce the results of measuring the observable $\hat{O}$ over the state $\rho_s$, i.e., to evaluate  $P_O=\mathrm{Tr}\{\hat{O}\rho_s\}$, one takes the following state
\begin{eqnarray}
\rho_s\rightarrow \rho_{s+a}=\frac{1}{C}\left(\sum\limits_{\forall c_j>0}|c_j|\rho_j|+\rangle\langle+|+
\sum\limits_{\forall c_k<0}|c_k|\rho_k|-\rangle\langle-|\right),
\label{cb1}
\end{eqnarray}
where $C=\sum\limits_{\forall j}|c_j|$ and the mutually orthogonal vectors $|\pm\rangle$ encode signs of the corresponding coefficients $c_j$. To evaluate $P_O$, it is necessary to perform a measurement over the  following observable
\[\hat{O}_{s+a}=C\hat{O}(|+\rangle\langle+|-|-\rangle\langle-|).\] 
Thus, one can reproduce the results of measurements of $\hat{O}$ over $\rho_s$ measuring $\hat{O}$ over the set $\rho_j$ conditioned on the result of measuring the ancilla. textcolor{blue}{For practical realization with single-mode states, the ancilla can be realized, for example, with an orthogonal polarization states of the additional mode  \cite{PhysRevA.105.052206,https://doi.org/10.1002/qute.202300060}. }

For fitting the single-photon state of an electromagnetic field mode according to Eq. (\ref{datapattern}), it is convenient to use phase-averaged coherent (PhAC) states 
\begin{eqnarray}
  \rho_j=\frac{1}{2\pi}\int\limits_{-\pi}^{\pi}d\phi \left||\alpha_j|e^{i\phi}\right\rangle\left\langle|\alpha_j|e^{i\phi}\right|= 
e^{-|\alpha_j|^2}\sum\limits_{n=0}^{\infty}\frac{|\alpha_j|^{2n}}{n!}|n\rangle\langle n|,
\label{cohev}  
\end{eqnarray}
where $|\alpha_j\rangle$ denote the coherent state with the amplitude $\alpha_j$, and $|n\rangle$ is the Fock state with $n$ photons. For the states diagonal in the Fock state basis the problem of fitting (\ref{datapattern}) is especially simple. For example, fitting the single photon state with just five phase-averaged coherent states with the amplitudes between 0 and 1 is sufficient for reaching the fidelity exceeding $99.9\%$  \cite{PhysRevA.105.052206,https://doi.org/10.1002/qute.202300060}. Notice that using more states for the representation gives one more freedom when finding $c_j$. As  will be seen below, it might better to build a representation with smaller  coefficients $c_j$ corresponding to smaller amplitudes for suppressing noise.  In the current work, we are using six  amplitudes in the interval from 0 to 1.45 for the representation. 

Using PhAC states allows modification of our method greatly simplifying experiments such as the HOM interference actually relying on measurements of the normally ordered correlation functions. Namely, the PhAC states  for measurement of such functions can be scaled. For creation $a_{1,2}^{\dagger}$
 and annihilation $a_{1,2}$ operators of the field modes $1$ and $2$, a general normally ordered  $n$-th order correlation function can be written as 
 \[G^{(n)}=\langle[a_1^{\dagger}]^m[a_2^{\dagger}]^{n-m}a_2^{n-m}a_1^m\rangle, \quad m\leq n.\] Let us take sets of PhAC states of the modes 1 and 2 with amplitudes $\{\alpha_j^{(1,2)}\}$ for building up representations of our quantum states. Then, if we scale all the states of the sets in the similar way,  $\alpha_j^{(1,2)}\mapsto x\alpha_j^{(1,2)}$, for any real $x$ it would be just $G^{(n)}\mapsto x^{2n}G^{(n)}$. Thus, it is possible to reproduce effects such as the HOM interference usually observed with the single-photon states, with macroscopic states and detectors operating far from the single-photon regime \cite{https://doi.org/10.1002/qute.202300060}. 

{After the quantum state has been conditionally prepared, it correctly reproduces the behavior of the genuine target quantum state for any kind of interference or detection scheme (for scaled conditional state preparation – only for measurement of normal-ordered field moments). Thus, such conditionally prepared state can be used to probe quantum setups and demonstrate their correct functioning, including the HOM interference setup discussed here.}

 \section{Experimental setup}
 
The schematic outline of the experiment is shown in Fig.~\ref{fig1}(a). There are two sources of electromagnetic field pulses (denoted as $A_{1,2}$ in Fig.~\ref{fig1}(a), or two outputs of the waveform generator in  Fig.~\ref{fig1}(c)) carrying the coherent states with amplitudes $x\alpha_{j}$. The angle $\varphi$  in Fig.~\ref{fig1}(a) represents a random phase between the states produced by $A_{1,2}$. The amplitudes $\alpha_{j}$ are  chosen from the set $\{\alpha_j\}$, $j=1,2 \ldots J$ devised to represent with high fidelity the single-photon state according to Eq. (\ref{datapattern}) with the coefficients $\{c_j\}$. The generated states subjected to phase averaging and time-delay were mixed on the 50/50 beam-splitter (BS in Fig.~\ref{fig1}(a) or the hybrid $90^o$ coupler in Fig.~\ref{fig1}(c)). The outputs of the beam-splitter are directed to corresponding detectors ($D_{1,2}$ in Fig.~\ref{fig1}(a) or inputs of the oscilloscope Fig.~\ref{fig1}(c)) producing signals proportional to the energy of the impinging pulse.  {Notice that there are two additional oscilloscope channels shown Fig.~\ref{fig1}(c). They are for monitoring signals at the inputs of the hybrid coupler and to calibrate the gain of the generator outputs taking into account various losses in cables and power dividers. }

Ideally, one needs to sample the amplitudes for each source according to the distribution $p_j=|c_j|/C$. However, assuming the limits of high number of samples, one can just perform measurement for a set of all possible pairs of amplitudes produced by the sources and sum the measurement results with weights equal to the products of corresponding coefficients $c_j$ for each source.  The PhAC states were obtained by averaging the phase for each state pair; the results were summed up with the a number of phases spanning a full circle, $[-\pi,\pi]$.  Also,  for each state pair the number of delays  were implemented for reproducing the typical dip of the correlations arising from the variation of the pulses' overlap. 

Fig.~\ref{fig1}(b) shows the photograph of the the experimental setup. Fig.~\ref{fig1}(c) presents the block-scheme of the setup. It was realized in the following way . The Keysight 81150A dual-channel arbitrary pulse generator formed two rectangular radio pulses at a frequency of 120 MHz with specified parameters. Each channel also contained internal noise of the device. The signals were fed to a specially made device consisting of a dual-channel 90° hybrid coupler, Mini-circuit QCV-151+ and two dual-channel 0° power dividers Mini-circuit ADP-2-1+. From the device output, the signals were fed to a Keysight DSOX3104T four-channel digital storage oscilloscope, which digitized and recorded the original and converted signals into the internal memory. The entire setup was controlled by a PC program implemented in Matlab. The program provided a cycle of signal formation and processing for all values of the relative delay of radio pulse envelopes (several thousand), amplitudes ($7\times7$ combinations for testing HOM interference of PhAC pairs and $6\times6$ for the actual HOM experiment) and the relative initial phase of radio pulses (16 values; it was found that more dense discretizations only marginally improve the results). The cycle duration was more than a day.

\begin{figure}
    \centering
    \includegraphics[width=0.99\linewidth]{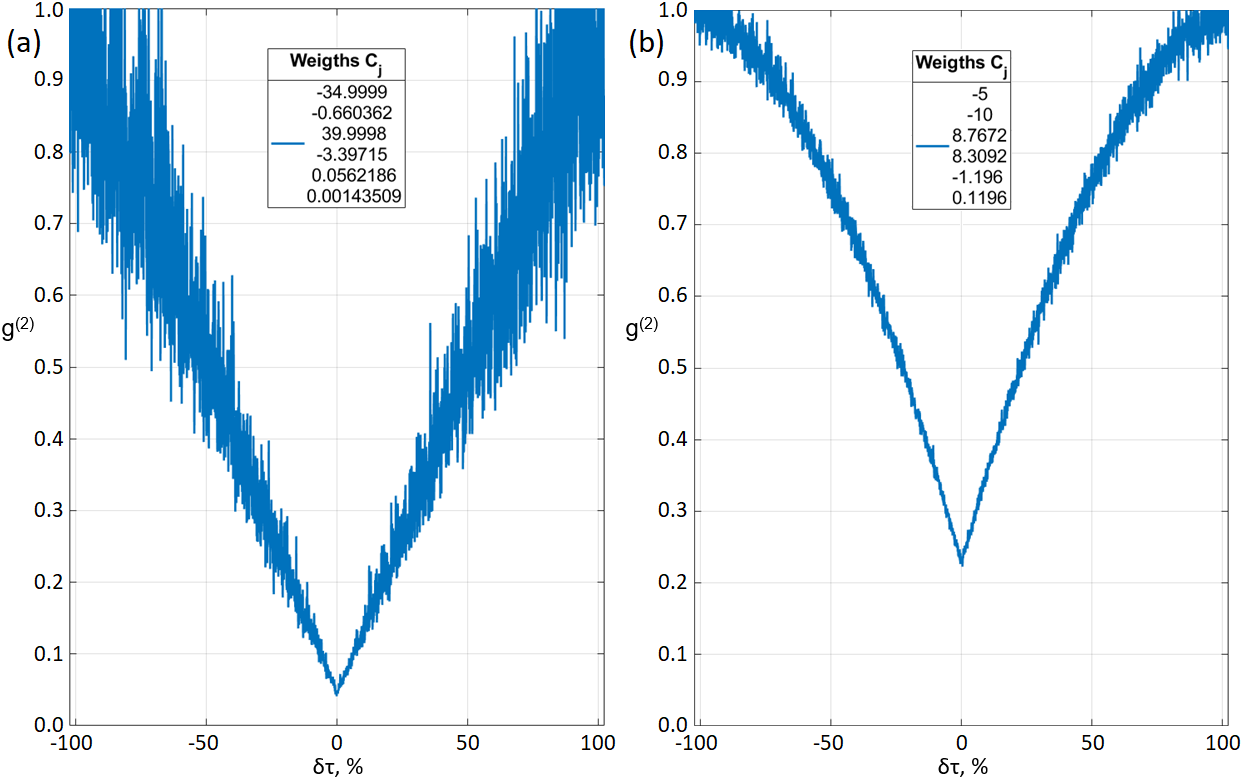}
    \caption{ (a,b) The HOM dip for the set of amplitudes shown for Fig.~\ref{fig2}(d) and the coefficients $c_j$ shown in the corresponding insets.}
    \label{fig3}
\end{figure}

 \section{Results}

First of all, to verify functioning of the setup, correlation measurements were performed for pairs of PhAC states with amplitudes from the set chosen for the conditional build-up of the single-photon states. For the similar rectangular pulses of  width $\tau$ carrying PhAC states interfering on the 50/50 beam-splitter (hybrid coupler), and $a_{1,2}$ being modes impinging on the detectors 1 and 2, the shape of the HOM dip is described by the following equation
\begin{eqnarray}
\nonumber
g^{(2)}_{12}(\delta\tau)=\frac{\langle a_1^{\dagger}(t)a_2^{\dagger}(t+\delta\tau)a_2(t+\delta\tau)a_1(t)\rangle}{\langle a_1^{\dagger}(t)a_1(t)\rangle\langle a_2^{\dagger}(t+\delta\tau)a_2(t+\delta\tau)\rangle}=\\
1-\frac{f^2}{2}, 
\quad f=\Biggl\{ \begin{matrix}
1-\frac{|\delta\tau|}{\tau},  -\tau\le\delta\tau\le \tau, \\
0, \quad |\delta\tau|>|\tau|.
\end{matrix}
\label{over1}    
\end{eqnarray}
where $f$ is the pulse overlap and $\delta\tau$ is the respective time-shift of the pulses impinging on the beam-splitter.  The shape given by Eq. (\ref{over1}) can be seen in Fig.~\ref{fig2}(c). In the absence of noise, the shape (\ref{over1}) should be observed for an arbitrary non-zero amplitude $\alpha_j$. The pulses actually implemented and observed in the experiments can be seen in Fig.~\ref{fig2}(a). The shape of pulses after leaving the beam-splitter can be seen in Fig.~\ref{fig2}(b) (notice that for demonstration of the HOM interference much longer pulses were used; namely, 82 periods per pulse were taken). One can see that there is noise (the right inset in Fig.~\ref{fig2}(d) shows the signal-to-noise ratio (SNR) for the average noise intensity and the amplitude of the pulse for different corresponding amplitudes $\alpha_j$), and also that the beam-splitting adds some noise and distorts pulse shapes even in the case of a supposedly complete  overlap. Thus, it is to be expected that the interference would be rather strongly impacted by the noise at lower amplitudes,  
and that the dip would not reach the value of 0.5 theoretically predicted for PhAC states. Indeed, one can see this in Fig.~\ref{fig2}(d), where HOM dips are shown with the states of different amplitudes (the respective amplitudes are shown in the left inset in  Fig.~\ref{fig2}(d)). {The noise causing interference degradation  stems  from the thermal noise of the generator, hybrid divider, and input circuits of the oscilloscope. Since we used a high-quality generator in the linear gain range, the power spectral density of noise does not change with a change in the output signal amplitude.  Therefore, for weaker signals, the relative contribution of the noise is larger. That spoils the interference effect.} For the results shown in  Fig.~\ref{fig2}(d), sets of 7 equal amplitude pulse pairs with 16 values of respective phase were taken.

{Now let us demonstrate how the quantum features of the Hong-Ou-Mandel interference can be captured with our set-up.} For  noiseless single-photon input one expects the HOM dip to be \cite{PhysRevA.105.052206,https://doi.org/10.1002/qute.202300060}
\begin{eqnarray}
g^{(2)}_{12}(\delta\tau) \propto
1-f^2.
\label{over2}    
\end{eqnarray}
{Specifically, for the ideal interference one expects a complete anti-correlation of signals recorded on the detectors. }
As  can be seen in  Fig.~\ref{fig3}, the respective dip extends far below the classically allowed value of 0.5. {But the complete anti-correlation, i.e., }zero value prophesied by Eq. (\ref{over2}) was not reached. As expected, noise influences also the HOM interference of conditionally built single-photon states. 
There is a possibility to mitigate the noise influence by building the state representation (\ref{datapattern}) with smaller coefficients $c_j$ corresponding to lower-amplitude states with relatively higher noise.

Fig.~\ref{fig3} shows examples of the HOM dips for conditionally built scaled single-photon inputs for two different sets of coefficients $\{c_j\}$. For both cases (just like for Fig.~\ref{fig2}(d) apart from the first amplitude) the set of PhAC with the following {relative} amplitudes was used: 
\begin{equation}
\alpha_{1 \ldots 6}=0.05, 0.1, 0.2, 0.4, 0.8, 1.45. 
\label{amps}
\end{equation}
Notice that due to the scaling feature mentioned in the Section II, only the ratios of the values (\ref{amps}) are important for the experiment.       We did not use the lowest amplitude $\alpha=0.01$ for the representation to reduce noise. For Fig.~\ref{fig3}(a), the upper limit  for evaluating the coefficients was set as $|c_j|\le 40$. This allowed to reach representation  fidelity exceeding $0.99$ and obtain the HOM dip reaching 0.05. However, large $c_j$ corresponding to the noisiest low-amplitude components lead to large noise. Suppressing the noisy components  by setting  $|c_1|\le 5$ and $|c_{2\ldots 6}|\le 10$ allowed to get much smoother HOM dip in Fig.~\ref{fig3}(b).  Thus, one needs to pay-off for the lower values of the coefficients with lower fidelity of the single-photon states' representation and larger coefficients corresponding to larger amplitude components (thus, larger respective weights of Fock states with photons numbers larger than one); for Fig.~\ref{fig3}(b) the fidelity of the representation falls below $0.95$  and the dip does not reach below 0.22. 

Overall, it has to be emphasized that in both cases for phase-averaged inputs interfering on the 50/50 beam-splitter we have obtained the HOM dips far below the classically allowed value of 0.5.

\section{Conclusions}

In this work, we have shown that the method of conditionally constructing quantum states from classical states {for demonstrating quantum effects} can be successfully implemented in the radio-frequency domain, using 120 MHz as a representative example. By employing ensembles of phase-averaged coherent states with amplitude ratios optimized to closely approximate a single-photon state, we observed Hong-Ou-Mandel interference dips in the normalized second-order correlation function that fall well below the classical threshold of 0.5. We also showed that the fidelity of the state construction can be tuned to mitigate the impact of experimental noise. These results provide experimental validation that conditional quantum state preparation is a practical and scalable approach for observing quantum interference effects in spectral regions where conventional quantum light sources are either unavailable or impractical.

Looking ahead, this technique opens the door {for demonstrating} other inherently quantum phenomena, such as violations of Bell inequalities, using radio waves. More broadly, it offers a flexible platform for {exploring and testing} quantum measurement schemes across a wide range of radio-frequency,  microwave and THz bands, well beyond the specific example of 120 MHz.

\section*{Acknowledgments} 

The authors (A.M., A.U. and P.K.) gratefully acknowledge financial support from Horizon Europe MSCA FLORIN Project 101086142. 
P.K. acknowledges financial support from
Research Council of Finland, decision PREIN no. 368653 and QuantERA project EXTRASENS, in Finland funded through Research Council of Finland, decision 361115. A.S. and K.B. acknowledge financial support from the Israel Ministry of Aliyah and Integration. T.S. acknowledges the RRF project Latvian Quantum Technologies Initiative no. 2.3.1.1.i.0/1/22/I/CFLA/001. The authors thank S.Vlasenko for the help with designing figures.

\bibliography{rhom2}

\end{document}